\documentclass[12pt]{article}

\usepackage{latexsym}

\usepackage{amssymb,amsfonts,amsmath}
\usepackage{indentfirst}

 \usepackage{bbm}

\parskip=\medskipamount

\newcommand {\cL}{{\cal L}}

\newcommand {\cN}{{\cal N}}

\def\a{\alpha}

\def\b{\beta}

\def\d{\delta}

\def\g{\gamma}
\def\G{\Gamma}

\def\j{\psi}

\def\l{\lambda}

\def\q{\theta}

\def\x{\xi}
\def\z{\zeta}

\def\F{\Phi}
\def\J{\Psi}
\def\L{\Lambda}

\def\S{\Sigma}

\def\rd{{\rm d}}
\def\ri{{\rm i}}

\newcommand{\ad}{{\dot{\alpha}}}                           
\newcommand{\bd}{{\dot{\beta}}}                            

\newcommand{\pa}{\partial}                           
\newcommand{\hf}{\frac12}

\newcommand{\be}{\begin{equation}}
\newcommand{\ee}{\end{equation}}
\newcommand{\bea}{\begin{eqnarray}}
\newcommand{\eea}{\end{eqnarray}}
\newcommand{\non}{\nonumber}
\newcommand{\1}{{\underline{1}}}
\newcommand{\2}{{\underline{2}}}

\newcommand{\bm}[1]{\mbox{\boldmath$#1$}}

\def\double #1{#1{\hbox{\kern-2pt $#1$}}}

\newif\ifdtup

\newcommand{\bsubeq}{\begin{subequations}}
\newcommand{\esubeq}{\end{subequations}}

\numberwithin{equation}{section}

\begin{document}
\begin{titlepage}
\begin{flushright}
November, 2017\\
\end{flushright}
\vspace{5mm}

\begin{center}
{\Large \bf The massless integer superspin multiplets revisited}\\ 
\end{center}

\begin{center}

{\bf Jessica Hutomo and Sergei M. Kuzenko
} \\
\vspace{5mm}

\footnotesize{
{\it School of Physics and Astrophysics M013, The University of Western Australia\\
35 Stirling Highway, Crawley W.A. 6009, Australia}}  
~\\
\texttt{20877155@student.uwa.edu.au,
sergei.kuzenko@uwa.edu.au }\\

\end{center}
\vspace{2mm}

\begin{flushright}
{\it Dedicated to Professor Ulf Lindstr\"om}\\
{\it on the occasion of his 70th birthday}
\end{flushright}

\begin{abstract}
\baselineskip=14pt
We propose a new off-shell formulation 
for the massless ${\cal N}=1$ supersymmetric 
multiplet of integer superspin $s$ in four dimensions, where $s =2,3,\dots$
(the $s=1$ case corresponds to the gravitino multiplet).
Its gauge freedom matches that of the superconformal 
superspin-$s$ multiplet described in arXiv:1701.00682.
The gauge-invariant action involves two compensating multiplets
in addition to the superconformal superspin-$s$ multiplet. 
Upon imposing a partial gauge fixing, this action reduces to the one describing the 
so-called longitudinal formulation for the  massless superspin-$s$ multiplet.
Our new model is shown to possess a dual realisation obtained by applying a superfield Legendre transformation. We present a non-conformal higher spin supercurrent multiplet associated with the new integer superspin theory.  
This fermionic supercurrent is shown to occur 
in the Fayet-Sohnius model for a massive ${\cal N}=2$ hypermultiplet. 
We also give a new off-shell realisation for the massless gravitino multiplet. 
\end{abstract}

\vfill

\vfill
\end{titlepage}

\newpage
\renewcommand{\thefootnote}{\arabic{footnote}}
\setcounter{footnote}{0}

\tableofcontents{}
\vspace{1cm}
\bigskip\hrule

\allowdisplaybreaks


\section{Introduction}
\setcounter{equation}{0}

In $\cN=1$  supersymmetric field theory in four dimensions, 
a massless multiplet of (half) integer superspin $\hat s >0$ describes two ordinary 
massless fields of spin $\hat s$ and $\hat s + \hf$. Such a supermultiplet is often denoted
$(\hat s, \hat s +\hf)$. The three lowest superspin values, 
$\hat s=\hf, 1$ and $ \frac{3}{2}$, 
correspond to the vector, gravitino  and supergravity multiplets, respectively.  
It follows from first principles that the sum of two actions 
for free massless spin-$\hat s$ and spin-$(\hat s+\hf)$ fields 
should possess an on-shell supersymmetry. This means that there is no 
problem of constructing on-shell massless higher superspin multiplets, 
with $\hat s > \frac{3}{2}$, 
for it is only necessary to work out the structure of supersymmetry transformations.
The latter task was completed first by Curtright  \cite{ Curtright} who made use of the 
(Fang-)Fronsdal actions \cite{Fronsdal,FF}, and soon after by Vasiliev \cite{Vasiliev}
who employed his frame-like reformulation of the (Fang-)Fronsdal models pioneered in 
\cite{Vasiliev}.  
Applications of the on-shell higher spin supermultiplets 
presented in \cite{ Curtright,Vasiliev} are 
rather limited. In particular, they do not allow one to construct supermultiplets
containing conserved higher spin currents that have to be off-shell, like the so-called
supercurrent multiplet \cite{FZ} containing the 
energy-momentum tensor and the supersymmetry current.
To obtain such higher spin supercurrents, 
off-shell realisations  for the massless higher superspin multiplets are required, 
and these are nontrivial to construct.\footnote{Early attempts to construct 
such off-shell realisations \cite{BO, BO2} were unsuccessful, as was explained in 
detail in \cite{KS94}.}

The problem of constructing gauge off-shell formulations for the  massless 
higher superspin multiplets was solved   in the early 1990s 
in the case of Poincar\'e supersymmetry \cite{KSP,KS}.\footnote{The results obtained
in  \cite{KSP,KS} are reviewed in \cite{Ideas}.}
For each superspin $\hat s > \frac{3}{2} $, 
half-integer \cite{KSP} and integer \cite{KS},
these publications provided two dually equivalent off-shell
actions formulated in ${\cal N }= 1$ Minkowski superspace.
At the component level, each of the
two superspin-$\hat s$ actions \cite{KSP,KS} reduces, 
{\it upon} imposing a Wess-Zumino-type
gauge and eliminating the auxiliary fields,
to a sum of the spin-$\hat s$ and  spin-$(\hat s+\hf)$ actions \cite{Fronsdal,FF}.
The massless higher superspin theories  of \cite{KSP,KS} were generalised 
to the case of anti-de Sitter supersymmetry in \cite{KS94}.

The non-supersymmetric higher spin theories of  \cite{Fronsdal,FF}
and their supersymmetric counterparts of half-integer superspin \cite{KSP}
share one common feature.  For each of them, 
the action is formulated in terms of a (super)conformal gauge (super)field 
coupled to certain compensators. Such a description does not yet exist for 
the massless supermultiplets of integer superspin $\hat s \geq 2$.  
One of the goals of this paper is to provide such a formulation
by properly generalising the off-shell supersymmetric actions 
given in \cite{KS}. We now make these points more precise.

Given an integer $s\geq 2$, the conformal spin-$s$ field \cite{FT,FL}
is described by a real potential\footnote{All tensor (super)fields encountered in this paper are completely symmetric 
with respect to their undotted spinor indices, and separately, 
with respect to their dotted indices. We use the notation
$V_{\a(s)\ad(t)} := V_{\a_1 \cdots \a_s \ad_1 \cdots \ad_t} 
= V_{(\a_1 \cdots \a_s)(\ad_1 \cdots \ad_t)} $
and
$V^{\a(s) \ad(t)} U_{\a(s)\ad(t)} := V^{\a_1 \cdots \a_s \ad_1 \cdots \ad_t} 
U_{\a_1 \cdots \a_s \ad_1 \cdots \ad_t}$.
Parentheses denote symmetrisation of indices; 
the undotted and dotted spinor indices are symmetrised independently. 
Indices sandwiched between vertical  bars 
(for instance,  $|\g|$) are not subject to symmetrisation. }
$h_{\a_1 \dots \a_{s} \ad_1 \dots \ad_{s} } 
=h_{(\a_1 \dots \a_{s}) (\ad_1 \dots \ad_{s} )} 
\equiv h_{\a(s) \ad(s)}$
with the gauge freedom
\begin{subequations}
\bea
 \d h_{\a_1 \dots \a_{s} \ad_1 \dots \ad_{s} } 
&=& \pa_{(\a_1 (\ad_1} \l_{\a_2\dots \a_{s}) \ad_2 \dots \ad_{s})}~,
\eea
for an arbitrary real gauge parameter 
$\l_{\a_1 \dots \a_{s-1}  \ad_1 \dots \ad_{s-1}}
=\l_{(\a_1 \dots \a_{s-1} )( \ad_1 \dots \ad_{s-1})} \equiv \l_{\a(s-1) \ad(s-1)}$. 
In addition to the gauge field $h_{\a(s) \ad(s)}$,
the massless spin-$s$ action \cite{Fronsdal} also involves  a real compensator
$h_{\a(s-2) \ad(s-2)}$ with the gauge transformation\footnote{For a review  
of the (Fang-)Fronsdal models \cite{Fronsdal,FF} in the two-component spinor notation 
 used in this paper, see e.g. \cite{Ideas}.}
\bea
 \d h_{\a_1 \dots \a_{s-2} \ad_1 \dots \ad_{s-2} } 
&=& \pa^{\b \bd} \l_{\b \a_1\dots \a_{s-2} \bd \ad_1 \dots \ad_{s-2}}~.
\eea
\end{subequations}
In the fermionic case, the conformal spin-$(s+\hf)$ field \cite{FT,FL}
is described by a potential
$\j_{\a(s+1) \ad(s)}$ and its conjugate $\bar \j_{\a(s) \ad(s+1)}$
with the gauge freedom
\begin{subequations}
\bea
 \d \j_{\a_1 \dots \a_{s+1} \ad_1 \dots \ad_{s} } 
&=& \pa_{(\a_1 (\ad_1} \x_{\a_2\dots \a_{s+1}) \ad_2 \dots \ad_{s})}~,
\eea
for an arbitrary gauge parameter 
$\x_{\a(s) \ad(s-1)}$.
In addition to the gauge fields $ \j_{\a(s+1) \ad(s)}$ and $\bar \j_{\a(s) \ad(s+1)}$, 
the massless spin-$(s+\hf)$ action \cite{FF} also involves two 
compensators $ \j_{\a(s-1) \ad(s)}$ and $ \j_{\a(s-1) \ad(s-2)}$
and their conjugates, with the the following gauge transformations
\bea
 \d \j_{\a_1 \dots \a_{s-1} \ad_1 \dots \ad_{s} } 
&=& \pa^\b{}_{ (\ad_1} \x_{\b\a_1\dots \a_{s-1} \ad_2 \dots \ad_{s})}~, \\
 \d \j_{\a_1 \dots \a_{s-1} \ad_1 \dots \ad_{s-2} } 
&=& \pa^{\b \bd} \x_{\b \a_1\dots \a_{s-1} \bd \ad_1 \dots \ad_{s-2}}~.
\eea
\end{subequations}

We now recall the structure of the off-shell higher spin supermultiplets.
Given a half-integer superspin $\hat s=s+\hf$, with 
$s=2,3, \dots$, the superconformal multiplet introduced in \cite{KMT} 
is described by a real  unconstrained prepotential 
$H_{\a(s)\ad(s)} $ 
possessing the gauge transformation law\footnote{In the $s=1$ case, 
the transformation law \eqref{1.3} corresponds to linearised 
conformal supergravity \cite{FZ2}. } 
\bea
 \d H_{\a_1 \dots \a_s \ad_1 \dots \ad_s} 
 = \bar D_{(\ad_1} \L_{\a_1 \dots \a_s \ad_2 \dots \ad_s )} 
- D_{(\a_1} \bar{\L}_{\a_2 \dots \a_s)\ad_1 \dots \ad_s} \ ,
\label{1.3}
\eea
with unconstrained gauge parameter $\L_{\a(s) \ad(s-1)}$. 
In addition to the gauge superfield $H_{\a(s)\ad(s)} $,
each of the massless superspin-$(s+\hf)$ actions constructed in \cite{KSP}
contains a compensating multiplet. In one case,
the compensating multiplet is described by a longitudinal linear superfield 
$G_{\a (s-1) \ad (s-1)}$ (and its conjugate $\bar G_{\a (s-1) \ad (s-1)}$)
constrained by 
\bea
{\bar D}_{ (\ad_1} \,G_{\a(s-1) \ad_2 \dots \ad_{s})}=0  \quad
&\Longrightarrow & \quad \bar D^2 G_{\a(s-1) \ad(s-1)}=0~,
\eea
with the gauge transformation
\bea
\d G_{\a_1 \dots \a_{s-1}\ad_1 \dots \ad_{s-1}} &= & - \hf \bar D_{(\ad_1} 
\bar D^{|\bd|} D^\b \L_{\b\a_1 \dots \a_{s-1} \ad_2 \dots \ad_{s-1}) \bd} \non\\
&&+ \ri (s-1) \bar D_{(\ad_1} \pa^{\b |\bd|} 
\L_{\b \a_1 \dots \a_{s-1} \ad_2 \dots \ad_{s-1} ) \bd} \ . 
\eea
In the other formulation, the compensating multiplet is described by 
a transverse linear superfield $\G_{\a (s-1) \ad (s-1)} $ 
(and its conjugate $\bar \G_{\a (s-1) \ad (s-1)} $)  constrained by 
\bea
{\bar D}^\bd \,\G_{\a(s-1) \bd \ad(s-2)} =  0 \quad
&\Longrightarrow & \quad \bar D^2 \G_{\a(s-1) \ad(s-1)}=0~,
\eea
 with the gauge transformation
\bea
\d_\L \G_{\a_1 \dots \a_{s-1}\ad_1 \dots \ad_{s-1}} &= &
-\frac{1}{4} \bar D^\bd D^2 \bar{\L}_{\a_1 \dots \a_{s-1}\bd \ad_1 
\dots \ad_{s-1}} ~.
\eea

Finally, in the case of an integer superspin $\hat s=s$, with 
$s=2,3, \dots$, the superconformal multiplet introduced in  \cite{KMT} 
 is described by an unconstrained prepotential $\J_{\a(s)\ad(s-1)} $ 
and its complex conjugate with the gauge transformation given by eq. \eqref{2.5a}
below, with unconstrained gauge parameters ${\frak V}_{\a(s-1) \ad(s-1)}$ 
and $\z_{\a(s) \ad(s-2)}$. The prepotential  $\J_{\a(s)\ad(s-1)} $ 
naturally occurs in the longitudinal formulation for the massless 
superspin-$s$ multiplet \cite{KS}. However, the gauge transformation of
$\J_{\a(s)\ad(s-1)} $ given in \cite{KS} differs from eq. \eqref{2.5a}. 
The difference is that the parameter ${\frak V}_{\a(s-1) \ad(s-1)}$ in \cite{KS} is not unconstrained, but instead is given by \eqref{2.99}.
In this paper we propose a new off-shell formulation for the massless higher integer superspin multiplet with the following properties: (i) the gauge freedom of 
$\J_{\a(s)\ad(s-1)} $ is given by \eqref{2.5a}; and (ii) the longitudinal formulation 
of \cite{KS} emerges upon imposing a gauge condition. 

This paper is organised as follows. In section 2 we present the new formulation 
for the massless superspin-$s$ multiplet. Its dual version is described in section 3.
In section 4 we introduce non-conformal higher spin supercurrents 
 associated  with the gauge massless superspin-$s$ multiplets.
 Section 5 is devoted to computing the higher spin supercurrents 
 that originate in the massive $\cN=2$ hypermultiplet model.
 Concluding comments are given in section 6, including a brief discussion of the 
 off-shell models for the massless gravitino multiplet. 


\section{New formulation}\label{section2}

Given a positive integer $s \geq 2$, we propose to describe the massless 
superspin-$s$ multiplet in terms of the following 
 superfield variables: 
(i) an unconstrained prepotential
 $\J_{\a(s)\ad(s-1)}  $ 
and its complex conjugate $\bar \J_{\a(s-1)\ad(s)}$; 
(ii) a real  superfield 
 $H_{\a(s-1)\ad(s-1)}  =\bar H_{\a(s-1)\ad(s-1)}  $; and (iii)
a complex superfield $\S_{\a(s-1) \ad (s-2) }$ 
and its conjugate $\bar \S_{\a (s-2) \ad(s-1)}$, 
where $\S_{\a(s-1) \ad (s-2) }$ is constrained 
to be transverse linear,\footnote{In
general, complex tensor superfields $\G_{\a(r) \ad(t)}$ and $G_{\a(r) \ad(t)}$ are called  transverse linear and longitudinal linear, respectively,
 if the constraints
$\bar D^\bd \G_{ \a(r) \bd \ad(t - 1) } = 0 $ and
$\bar D_{(\bd} G_{\a(r)\ad_1 \dots \ad_t )} = 0 $ are satisfied. 
The former constraint is defined for $t \neq 0$; it has to be replaced 
with the standard linear 
constraint,  $\bar D^2 \G_{ \a(r) } = 0 $, for $t=0$.
The latter constraint for $t = 0$ is the chirality condition
$\bar D_\bd G_{ \a(r) } = 0$.
}
\bea
 \bar D^\bd \S_{\a(s-1) \bd \ad(s-3)} =0~.
 \label{2.1}
 \eea
  In the $s=2$ case, 
 for which \eqref{2.1} is not defined, 
 $\S_{\a(2)} $ is instead constrained to be complex linear, 
 \bea
 \bar D^2 \S_{\a(2) } =0~.
 \label{2.2}
 \eea
 The constraint \eqref{2.1}, or its counterpart \eqref{2.2} for $s=2$, 
  can be solved in terms of a complex unconstrained
 prepotential $Z_{\a(s-1) \ad (s-1)}$ by the rule
 \bea
 \S_{\a(s-1) \ad (s-2)} = \bar D^\bd Z_{\a(s-1) (\bd \ad_1 \dots \ad_{s-2} )} ~.
 \label{2.3}
 \eea
 This prepotential is defined modulo 
 gauge transformations 
 \bea
 \d_\x Z_{\a(s-1) \ad (s-1)}=  \bar D^\bd \x_{\a(s-1) (\bd \ad_1 \dots \ad_{s-1} )} ~,
 \label{2.4}
 \eea
 with the gauge parameter  $\x_{\a(s-1) \ad (s) }$ being  unconstrained.

The gauge freedom of $\J_{\a_1 \dots \a_s \ad_1 \dots \ad_{s-1}} $ is
chosen to coincide with that of the superconformal superspin-$s$ multiplet 
\cite{KMT}, which is
\begin{subequations} \label{2.5}
\bea
 \d_{ {\frak V} ,\z} \J_{\a_1 \dots \a_s \ad_1 \dots \ad_{s-1}} 
 &=& \hf D_{(\a_1}  {\frak V}_{\a_2 \dots \a_s)\ad_1 \dots \ad_{s-1}}
+  \bar D_{(\ad_1} \z_{\a_1 \dots \a_s \ad_2 \dots \ad_{s-1} )}  ~ , \label{2.5a}
\eea
with unconstrained gauge parameters ${\frak V}_{\a(s-1) \ad(s-1)}$ 
and $\z_{\a(s) \ad(s-2)}$. 
The $\frak V$-transformation is defined to act on the superfields $H_{\a(s-1) \ad(s-1)}$
and $\S_{\a(s-1) \ad(s-2) }$ as follows
\bea
\d_{\frak V} H_{\a(s-1) \ad(s-1)}&=& {\frak V}_{\a(s-1) \ad(s-1)} +\bar {\frak V}_{\a(s-1) \ad(s-1)}
~, \label{2.5b}\\
\d_{\frak V} \S_{\a(s-1) \ad(s-2) }&=&  \bar D^\bd \bar {\frak V}_{\a(s-1) \bd \ad(s-2)}
\quad \Longrightarrow \quad \d_{\frak V} Z_{\a(s-1) \ad (s-1)}
=\bar  {\frak V}_{\a(s-1) \ad (s-1)}~.~~~
\label{2.5c}
\eea
\end{subequations}
The longitudinal linear superfield 
\bea
G_{\a_1 \dots \a_s \ad_1 \dots \ad_s} := 
\bar D_{(\ad_1} \J_{\a_1 \dots \a_s \ad_2 \dots \ad_s)}~, 
\qquad \bar D_{(\ad_1} G_{\a_1 \dots \a_s \ad_2 \dots \ad_{s+1})}=0
\label{2.6}
\eea
is invariant under the $\z$-transformation \eqref{2.5a} 
and  varies under the $\frak V$-transformation as 
\bea
 \d_{ {\frak V} } G_{\a_1 \dots \a_s \ad_1 \dots \ad_{s}} 
 &=& \hf \bar D_{(\ad_1} D_{(\a_1}  {\frak V}_{\a_2 \dots \a_s)\ad_2 \dots \ad_{s})}~.
 \eea

It may be checked that the following action
\bea
S^{\|}_{(s)} &=&
\Big( - \frac{1}{2}\Big)^s  \int 
 \rd^4x \rd^2 \q  \rd^2 \bar \q
\,
\left\{ \frac{1}{8} H^{ \a (s-1) \ad (s-1) }  D^\b {\bar D}^2 D_\b 
H_{\a (s-1) \ad (s-1)} \right. \non \\
&&+ \frac{s}{s+1}H^{ \a(s-1) \ad(s-1) }
\Big( D^{\b}  {\bar D}^{\bd} G_{\b\a(s-1) \bd\ad(s-1) }
- {\bar D}^{\bd}  D^{\b} 
{\bar G}_{\b \a (s-1) \bd \ad (s-1) } \Big) \non \\
&&+ 2 \bar G^{ \a (s) \ad (s) } G_{ \a (s) \ad (s) } 
+ \frac{s}{s+1}\Big( G^{ \a (s) \ad (s) } G_{ \a (s) \ad (s) } 
+ \bar G^{ \a (s) \ad (s) }  \bar G_{ \a (s) \ad (s) } 
 \Big) \non \\
 &&+ \frac{s-1}{4s}H^{ \a(s-1) \ad(s-1) }
\Big( D_{\a_1} \bar D^2 \bar \S_{\a_2 \dots \a_{s-1}\ad(s-1)}
 - {\bar D}_{\ad_1}  D^2 \S_{\a(s-1) \ad_2 \dots \ad_{s-1} } \Big)  \non \\
&&+\frac{1}{s} \J^{\a(s) \ad(s-1)} \Big( 
D_{\a_1} \bar D_{\ad_1} -2\ri (s-1) \pa_{\a_1 \ad_1} \Big)
\S_{\a_2 \dots \a_s \ad_2 \dots \ad_{s-1} }\non  \\
&&+\frac{1}{s} \bar \J^{\a(s-1) \ad(s)} \Big( 
 \bar D_{\ad_1} D_{\a_1}-2\ri (s-1) \pa_{\a_1 \ad_1} \Big)
\bar \S_{\a_2 \dots \a_{s-1} \ad_2 \dots \ad_{s} }\non \\
&&+ \frac{s-1}{8s} \Big( \S^{\a(s-1) \ad(s-2) } D^2 \S_{\a(s-1) \ad(s-2)} 
- \bar \S^{\a(s-2) \ad(s-1) }\bar D^2 \bar \S_{\a(s-2) \ad(s-1)} \Big)
\non \\
&& \left.
- \frac{1}{s^2}\bar \S^{\a(s-2) \ad(s-2)\bd } \Big( \hf (s^2 +1) D^\b \bar D_\bd 
+\ri  {(s-1)^2} \pa^\b{}_\bd \Big) \S_{\b \a(s-2) \ad(s-2)} \right\}
\label{action}
\eea
is invariant under the gauge transformations \eqref{2.5}.
By construction, the action is also invariant under \eqref{2.4}.

The $\frak V$-gauge freedom \eqref{2.5} may be used to impose the condition 
\bea
 \S_{\a(s-1) \ad(s-2) }=0~.
\label{2.9}
\eea
In this gauge, the action \eqref{action} reduces to 
that describing the longitudinal formulation for the massless superspin-$s$ multiplet  \cite{KS}. The gauge condition \eqref{2.9} does not 
fix completely the $\frak V$-gauge freedom. The residual gauge transformations  
are generated by 
\bea
{\frak V}_{\a(s-1) \ad(s-1)} = D^\b L_{(\b \a_1 \dots \a_{s-1}) \ad(s-1)}~,
\label{2.99}
\eea
with the parameter $L_{\a(s) \ad(s-1)}$ being an unconstrained superfield. 
With this expression for ${\frak V}_{\a(s-1) \ad(s-1)}$, the gauge transformations \eqref{2.5a}  and \eqref{2.5b} coincide with those given in \cite{KS}.
Our consideration implies that the action \eqref{action} indeed provides an off-shell formulation for the massless superspin-$s$ multiplet . 

Instead of choosing the condition \eqref{2.9},
one can impose an alternative gauge fixing 
\bea
H_{\a(s-1) \ad(s-1)} =0~.
\label{2.10}
\eea
In accordance with \eqref{2.5b}, in this gauge
the residual gauge freedom is described by 
\bea
{\frak V}_{\a(s-1) \ad(s-1)} = \ri {\frak R}_{\a(s-1) \ad(s-1)}~, \qquad 
\bar{\frak R}_{\a(s-1) \ad(s-1)}={\frak R}_{\a(s-1) \ad(s-1)}~.
\eea

The action \eqref{action} includes a single term which involves the `naked' 
gauge field $\J_{\a(s)\ad(s-1)} $ and not the field strength $G_{\a(s)\ad(s)} $, 
the latter being 
defined by \eqref{2.6} and invariant under the $\z$-transformation \eqref{2.5a}.
This is actually a BF term, for it can be written in two different forms
\bea
\frac{1}{s}  \int  \rd^4x \rd^2 \q  \rd^2 \bar \q
 \,
 \J^{\a(s) \ad(s-1)} \Big( 
D_{\a_1} \bar D_{\ad_1} &-&2\ri (s-1) \pa_{\a_1 \ad_1} \Big)
\S_{\a_2 \dots \a_s \ad_2 \dots \ad_{s-1} } \non \\
=- \frac{1}{s+1}  \int 
 \rd^4x \rd^2 \q  \rd^2 \bar \q
\,
 G^{\a(s) \ad(s)} \Big( \bar D_{\ad_1}D_{\a_1}  
&+&2\ri (s+1) \pa_{\a_1 \ad_1} \Big)
Z_{\a_2 \dots \a_s \ad_2 \dots \ad_{s} }~.
\label{2.12}
\eea
The former makes the gauge symmetry \eqref{2.4} manifestly realised, 
while the latter
turns the $\z$-transformation \eqref{2.5a} into a manifest symmetry.

Making use of \eqref{2.12} leads to a different representation for the action
\eqref{action}. It is 
\bea
S^{\|}_{(s)} &=&
\Big( - \frac{1}{2}\Big)^s  \int 
 \rd^4x \rd^2 \q  \rd^2 \bar \q
\,
\left\{ \frac{1}{8} H^{ \a (s-1) \ad (s-1) }  D^\b {\bar D}^2 D_\b 
H_{\a (s-1) \ad (s-1)} \right. \non \\
&&+ \frac{s}{s+1}H^{ \a(s-1) \ad(s-1) }
\Big( D^{\b}  {\bar D}^{\bd} G_{\b\a(s-1) \bd\ad(s-1) }
- {\bar D}^{\bd}  D^{\b} 
{\bar G}_{\b \a (s-1) \bd \ad (s-1) } \Big) \non \\
&&+ 2 \bar G^{ \a (s) \ad (s) } G_{ \a (s) \ad (s) } 
+ \frac{s}{s+1}\Big( G^{ \a (s) \ad (s) } G_{ \a (s) \ad (s) } 
+ \bar G^{ \a (s) \ad (s) }  \bar G_{ \a (s) \ad (s) } 
 \Big) \non \\
 &&+ \frac{s-1}{4s}H^{ \a(s-1) \ad(s-1) }
\Big( D_{\a_1} \bar D^2 \bar \S_{\a_2 \dots \a_{s-1}\ad(s-1)}
 - {\bar D}_{\ad_1}  D^2 \S_{\a(s-1) \ad_2 \dots \ad_{s-1} } \Big)  \non \\
&&
- \frac{1}{s+1}  G^{\a(s) \ad(s)} \Big( \bar D_{\ad_1}D_{\a_1}  
+2\ri (s+1) \pa_{\a_1 \ad_1} \Big)
Z_{\a_2 \dots \a_s \ad_2 \dots \ad_{s} }
\non  \\
&&
+ \frac{1}{s+1}  \bar G^{\a(s) \ad(s)} \Big(  D_{\a_1}\bar D_{\ad_1}  
+2\ri (s+1) \pa_{\a_1 \ad_1} \Big)
\bar Z_{\a_2 \dots \a_s \ad_2 \dots \ad_{s} }
\non \\
&&+ \frac{s-1}{8s} \Big( \S^{\a(s-1) \ad(s-2) } D^2 \S_{\a(s-1) \ad(s-2)} 
- \bar \S^{\a(s-2) \ad(s-1) }\bar D^2 \bar \S_{\a(s-2) \ad(s-1)} \Big)
\non \\
&& \left.
- \frac{1}{s^2}\bar \S^{\a(s-2) \ad(s-2)\bd } \Big( \hf (s^2+1)D^\b \bar D_\bd 
+\ri  {(s-1)^2} \pa^\b{}_\bd \Big) \S_{\b \a(s-2) \ad(s-2)} \right\}~.~~~
\label{action2}
\eea


\section{Dual formulation}

The theory with action \eqref{action2} possesses a dual formulation 
that can be obtained by applying the duality transformation introduced in \cite{KSP,KS}.
In general, it works as follows. Suppose we have a supersymmetric field 
theory formulated in terms of a longitudinal linear superfield
$G_{\a(t) \ad(s)} $ and its conjugate $\bar G_{\a(s) \ad(t)}$, 
and the action has the form
\bea
S[ G_{\a(t) \ad(s)} , \bar G_{\a(s) \ad(t)}]=
  \int  \rd^4x \rd^2 \q  \rd^2 \bar \q\,
  \cL\Big(G_{\a(t) \ad(s)} , \bar G_{\a(s) \ad(t)}\Big)~,
\label{2.14}  
\eea
where $\cL(G, \bar G)$ is an algebraic function of its arguments. 
We now associate with this theory  
a first-order model of the form
\bea
S_{\text{first-order} }=
  \int  \rd^4x \rd^2 \q  \rd^2 \bar \q\, \left\{
  \cL\Big(U_{\a(t) \ad(s)} , \bar U_{\a(s) \ad(t)}\Big)
  +  \Big( \G^{\a(t) \ad(s)} U_{\a(t) \ad(s)} 
  +{\rm c.c.} \Big)\right\}~,~~~
  \label{2.15}
  \eea
  where $U_{\a(t) \ad(s)} $ is a complex unconstrained superfield, 
  and the Lagrange multiplier $\G_{\a(t) \ad(s)} $ is transverse linear. 
  Varying $S_{\text{first-order} }$ with respect to the Lagrange multiplier
  gives $U_{\a(t) \ad(s)} = G_{\a(t) \ad(s)} $, and then  $S_{\text{first-order} }$
  reduces to the original action \eqref{2.14}.
On the other hand, we can consider the equation of motion for $U^{\a(t) \ad(s)} $, 
which is
\bea
\frac{\pa }{\pa U^{\a(t) \ad(s)} }  \cL\Big(U_{\b(t) \bd(s)} , \bar U_{\b(s) \bd(t)}\Big)
+ \G_{\a(t) \ad(s)} =0~.
\label{2.16}
\eea
we assume that \eqref{2.16} can be solved to express $U_{\b(t) \bd(s)} $
in terms of $\G_{\a(t) \ad(s)} $ and its conjugate. Plugging this solution 
back into \eqref{2.15} gives a dual action 
\bea
S_{\rm dual}[ \G_{\a(t) \ad(s)} , \bar \G_{\a(s) \ad(t)}]=
  \int  \rd^4x \rd^2 \q  \rd^2 \bar \q\,
  \cL_{\rm dual}\Big(\G_{\a(t) \ad(s)} , \bar \G_{\a(s) \ad(t)}\Big)~.
\eea

In the $t=s=0$ case, the above duality transformation 
coincides with the so-called complex linear--chiral duality \cite{GGRS}
which  plays a fundamental role in the context 
of off-shell supersymmetric sigma models with eight supercharges 
\cite{LR,GK}.

We now associate with our theory \eqref{action2} 
the following first-order action\footnote{The specific normalisation of the Lagrange multiplier in \eqref{action3} is chosen to match that of \cite{KS94,KS}.} 
\bea
S_{\text{first-order} }&=&S^{\|}_{(s)}[U, \bar U, H , Z, \bar Z]  \non \\
&&+ \Big( \frac{-1}{2} \Big)^s\int  \rd^4x \rd^2 \q  \rd^2 \bar \q\,
 \Big(\frac{2}{s+1} \G^{\a(s) \ad(s)} U_{\a(s) \ad(s)} 
  +{\rm c.c.} \Big)~,~~
\label{action3}
\eea
where $S^{\|}_{(s)}[U, \bar U, H , Z, \bar Z] $ is obtained from the action
\eqref{action2} by replacing $G_{\a(s) \ad(s)} $ with an unconstrained 
complex superfield $U_{\a(s) \ad(s)} $, and $\G_{\a(s) \ad(s)} $
is a transverse linear superfield, 
\bea
\bar D^\bd \G_{\a(s) \bd \ad_1 \dots \ad_{s-1}} =0 ~.
\eea
As discussed above, the first-order model introduced 
is equivalent to the original theory 
 \eqref{action2}. The action  \eqref{action3} is invariant under the gauge $\x$-transformation 
 \eqref{2.4} which acts on $U_{\a (s) \ad (s)}$ and
 $\G_{\a(s) \ad(s)}$ by the rule
\begin{subequations}\label{3.7}
 \bea
 \d_\x U_{\a (s) \ad (s)} &=&0~,\\
 \d_\x \G_{\a(s) \ad(s)} &=& \bar D^\bd \Big\{  \frac{s+1}{2(s+2)}
\bar D_{(\bd} D_{\a_1} \x_{\a_2 \dots \a_s \ad_1 \dots \ad_s) } 
+\ri (s+1)\pa_{\a_1 (\bd } \x_{\a_2 \dots \a_s \ad_1 \dots \ad_s) } \Big\}~.~~~
\label{3.7b}
\eea
\end{subequations}
The first-order action  \eqref{action3} is also invariant under the gauge $\frak V$-transformation \eqref{2.5b} and \eqref{2.5c}, which acts on $U_{\a (s) \ad (s)}$ and
$\G_{\a(s) \ad(s)} $ as
\begin{subequations}
\bea
 \d_{ {\frak V} } U_{\a (s) \ad (s)} 
 &=& \hf \bar D_{(\ad_1} D_{(\a_1}  {\frak V}_{\a_2 \dots \a_s)\ad_2 \dots \ad_{s})}~, \\
\d_{\frak V} \G_{\a(s) \ad(s)} &=&0~.
\eea 
\end{subequations}

Eliminating the auxiliary superfields  $U_{\a(s) \ad(s)} $ and  $\bar U_{\a(s) \ad(s)} $ 
from \eqref{action3} leads to 
\bea
S^{\perp}_{(s)} &=& - \Big( - \hf \Big)^s 
  \int  \rd^4x \rd^2 \q  \rd^2 \bar \q\,
\Bigg\{ - \frac{1}{8} H^{\a(s-1) \ad(s-1)} D^\b \bar D^2  D_\b H_{\a(s-1)\ad(s-1)} \non\\
&&+ \frac{1}{8} \frac{s^2}{(s+1)(2s+1)} [D^\b, \bar D^\bd] H^{\a(s-1)\ad(s-1)} 
[D_{(\b}, \bar D_{(\bd}] H_{\a(s-1))\ad(s-1))} \non\\
&&+\hf \frac{s^2}{s+1} \pa^{\b\bd} H^{\a(s-1)\ad(s-1)} \pa_{(\b(\bd} H_{\a(s-1))\ad(s-1))}
 \non\\
&&+ \frac{2 \ri s}{2s+1} H^{\a(s-1)\ad(s-1)} \pa^{\b\bd} 
\Big({\bm \G}_{\b\a(s-1)\bd\ad(s-1)} 
- \bar{\bm \G}_{\b\a(s-1)\bd\ad(s-1)}\Big) \non\\
&& 
+ \frac{2}{2s+1} \bar{\bm \G}^{ \a (s) \ad (s) } {\bm \G}_{ \a (s) \ad (s) } 
- \frac{s}{(s+1)(2s+1)} \Big({\bm \G}^{ \a (s) \ad (s) }  {\bm \G}_{ \a (s) \ad (s) } 
+ \bar{\bm \G}^{ \a (s) \ad (s) } \bar{\bm \G}_{ \a (s) \ad (s) }\Big) \non\\
 &&- \frac{s-1}{2(2s+1)} H^{ \a(s-1) \ad(s-1) }
\Big( D_{\a_1} \bar D^2 \bar \S_{\a_2 \dots \a_{s-1}\ad(s-1)}
 - {\bar D}_{\ad_1}  D^2 \S_{\a(s-1) \ad_2 \dots \ad_{s-1} } \Big)  \non \\
&&+ \frac{1}{2(2s+1)} H^{ \a(s-1) \ad(s-1) }
\Big(  D^2 {\bar D}_{\ad_1} \S_{\a(s-1) \ad_2 \dots \ad_{s-1} }
 - \bar D^2 D_{\a_1} \bar \S_{\a_2 \dots \a_{s-1}\ad(s-1)} \Big)  \non \\
&&- \ri \frac{(s-1)^2}{s(2s+1)} H^{ \a(s-1) \ad(s-1) }
 \pa_{\a_1 \ad_1} \Big( D^\b \S_{ \b \a_2 \dots \a_{s-1} \ad_2 \dots \ad_{s-1}}
+ \bar D^\bd \bar \S_{ \a_2 \dots \a_{s-1} \bd \ad_2 \dots \ad_{s-1}} \Big)  \non \\
&&- \frac{s-1}{8s} \Big( \S^{\a(s-1) \ad(s-2) } D^2 \S_{\a(s-1) \ad(s-2)} 
- \bar \S^{\a(s-2) \ad(s-1) }\bar D^2 \bar \S_{\a(s-2) \ad(s-1)} \Big)
\non \\
&&
+ \frac{1}{s^2}\bar \S^{\a(s-2) \ad(s-2)\bd } \Big( \hf (s^2 +1) D^\b \bar D_\bd 
+\ri  {(s-1)^2} \pa^\b{}_\bd \Big) \S_{\b \a(s-2) \ad(s-2)}
\Bigg\} ~,
\label{action4}
\eea
 where we have defined
\bea
{\bm \G}_{ \a (s) \ad (s) } = \G_{ \a (s) \ad (s) }
-\hf \bar D_{(\ad_1} D_{(\a_1} Z_{\a_2 \dots \a_s) \ad_2 \dots \ad_s) } 
-\ri (s+1)\pa_{(\a_1 (\ad_1 } Z_{\a_2 \dots \a_s) \ad_2 \dots \ad_s) } ~.~~~
\label{shifted}
\eea
We point out that ${\bm \G}_{ \a (s) \ad (s) } $ is invariant under 
the gauge transformations \eqref{2.4} and \eqref{3.7b}.

In accordance with \eqref{2.5c}, the gauge $\frak V$-freedom may be used to
impose the condition 
\bea
Z_{\a(s-1) \ad(s-1)} =0~.
\label{3.11}
\eea
In this gauge the action \eqref{action4} reduces to the one defining 
the transverse formulation for the massless superspin-$s$ multiplet \cite{KS}.
The gauge condition \eqref{3.11} is preserved by residual local 
$\frak V$- and $\x$-transformations of the form 
\bea
  \bar D^\bd \x_{\a(s-1) \bd \ad (s-1 )}  +
  \bar {\frak V}_{\a(s-1)\ad (s-1 )} =0~.
\eea
 Making use of the parametrisation \eqref{2.99}, the residual gauge freedom is
\begin{subequations}
\bea
\d H_{\a(s-1)\ad(s-1)} &=& D^\b L_{\b \a(s-1) \ad(s-1)} - \bar D^\bd \bar{L}_{\a(s-1)\bd\ad(s-1)} \ ,\\
\d \G_{\a(s) \ad(s)} &=& \frac{s+1}{2(s+2)} \bar D^\bd \Big\{ 
\bar D_{(\bd} D_{(\a_1} 
+ 2\ri (s+2) \pa_{(\a_1(\bd}\Big\}
 \bar{L}_{\a_2 \dots \a_{s})\ad_1 \dots \ad_{s})} \ ,
\eea
\end{subequations}
which is exactly the gauge symmetry of the transverse formulation for the massless superspin-$s$ multiplet \cite{KS}.


\section{Higher spin supercurrent multiplets}

We now make use of the new gauge formulation \eqref{action}, 
or  equivalently \eqref{action2}, 
 for the integer superspin-$s$ multiplet
to derive  non-conformal higher spin supercurrents.

Let us couple the prepotentials 
$H_{ \a (s-1) \ad (s-1) } $, $Z_{ \a (s-1) \ad (s-1) }$ and $\Psi_{ \a (s) \ad (s-1) } $ to external sources
\bea
S^{(s)}_{\rm source} &=& \int \rd^4x \rd^2 \q  \rd^2 \bar \q \, \Big\{ 
\Psi^{ \a (s) \ad (s-1) } J_{ \a (s) \ad (s-1) }
-\bar \Psi^{ \a (s-1) \ad (s) } \bar J_{ \a (s-1) \ad (s) }
\non \\
&&+H^{ \a (s-1) \ad (s-1) } S_{ \a (s-1) \ad (s-1) } \non \\
&&+ Z^{ \a (s-1) \ad (s-1) } T_{ \a (s-1) \ad (s-1) } 
+ \bar Z^{ \a (s-1) \ad (s-1) } \bar T_{ \a (s-1) \ad (s-1) }
 \Big\}~.
\label{4.1}
\eea
In order for $S^{(s)}_{\rm source}$ to be invariant under the $\z$-transformation 
in \eqref{2.5a}, the source  $J_{ \a (s) \ad (s-1) }$ must satisfy
\bea
\bar D^\bd J_{\a(s) \bd \ad(s-2)} =0 \quad \Longleftrightarrow \quad
D^\b \bar J_{\b \a(s-2)  \ad(s)} =0 ~.
\label{4.2a}
\eea
Next, in order for $S^{(s)}_{\rm source}$ to be invariant under the transformation \eqref{2.4},
the source
$T_{ \a (s-1) \ad (s-1) }$ must satisfy
\bea
\bar D_{(\ad_1} T_{\a(s-1) \ad_2 \dots \ad_{s})} =0
 \quad \Longleftrightarrow \quad
 D_{(\a_1} \bar T_{\a_2 \dots \a_{s})  \ad (s-1)} =0~.
\label{4.2b}
\eea
We see that  the superfields $J_{ \a (s) \ad (s-1) }$ and $T_{ \a (s-1) \ad (s-1) } $ are transverse linear and longitudinal linear, respectively.
Finally, requiring $S^{(s)}_{\rm source}$ to be invariant under the 
$\frak V$-transformation 
\eqref{2.5} gives the following conservation equation
\begin{subequations} \label{3.4}
\bea
-\hf D^\b J_{\b \a(s-1) \ad(s-1)} 
+S_{\a(s-1) \ad(s-1)} + \bar T_{\a(s-1) \ad(s-1)} =0
\label{4.2c}
\eea
and its conjugate
\bea
\hf \bar D^\bd \bar J_{ \a(s-1) \bd \ad(s-1)} 
+S_{\a(s-1) \ad(s-1)} + T_{\a(s-1) \ad(s-1)} =0~. 
\label{3.4b}
\eea
\end{subequations}

As a consequence of  \eqref{4.2b}, from \eqref{4.2c} we deduce
\bea
\frac{1}{4} D^2 J_{ \a(s) \ad(s-1)} + D_{(\a_1} S_{\a_2 \dots \a_s) \ad(s-1) } =0~.
\label{3.5}
\eea
The equations \eqref{4.2a} and \eqref{3.5} describe the conserved 
current supermultiplet which corresponds to our theory in the gauge \eqref{2.9}.

Taking the sum of \eqref{4.2c} and \eqref{3.4b}
leads to
\bea
\hf D^\b J_{\b \a(s-1) \ad(s-1)} 
+\hf \bar D^\bd \bar J_{\a(s-1) \bd \ad(s-1)}
+ T_{\a(s-1) \ad(s-1)}-\bar T_{\a(s-1) \ad(s-1)} =0~. 
\label{4.3}
\eea
The equations \eqref{4.2a}, \eqref{4.2b} and \eqref{4.3} describe the conserved 
current supermultiplet which corresponds to our theory in the gauge \eqref{2.10}.
As a consequence of \eqref{4.2b}, the conservation equation \eqref{4.3} 
implies
\bea
\hf D_{(\a_1} \left\{D^{|\b|} J_{\a_2 \dots \a_s ) \b\ad(s-1)} 
+ \bar D^\bd \bar J_{\a_2 \dots \a_s ) \bd \ad(s-1)}\right\}
+D_{(\a_1} T_{\a_2 \dots \a_s ) \ad(s-1)} =0~. 
\label{4.4}
\eea

As in \cite{HK}, it is useful to introduce 
auxiliary complex variables $\z^\a \in {\mathbb C}^2$ and their conjugates 
$\bar \z^\ad$. Given a tensor superfield $U_{\a(p) \ad(q)}$, we associate with it 
the following  field on ${\mathbb C}^2$ 
\bea
U_{(p,q)} (\z, \bar \z):= \z^{\a_1} \dots \z^{\a_p} \bar \z^{\ad_1} \dots \bar \z^{\ad_q}
U_{\a_1 \dots \a_p \ad_1 \dots \ad_q}~,
\eea
which is homogeneous of degree $(p,q)$ in the variables $\z^\a$ and $\bar \z^\ad$.
We introduce operators that  increase the degree 
of homogeneity in the variables $\z^\a$ and $\bar \z^\ad$, 
\begin{subequations}
\bea
{D}_{(1,0)} &:=& \z^\a D_\a~, \qquad {D}_{(1,0)}^2=0~, \\
{\bar D}_{(0,1)} &:=& \bar \z^\ad \bar D_\ad~, 
\qquad {\bar D}_{(0,1)}^2 =0~, \\
{\pa}_{(1,1)} &:=& 2\ri \z^\a \bar \z^\ad \pa_{\a\ad}
= -\big\{ {D}_{(1,0)} , \bar {D}_{(1,0)} \big\}
~.
\eea
\end{subequations}
We also introduce two {\it nilpotent} operators that decrease the degree 
of homogeneity in the variables $\z^\a$ and $\bar \z^\ad$, specifically
\begin{subequations}
\bea
D_{(-1,0)} &:=& D^\a \frac{\pa}{\pa \z^\a}~, \qquad D_{(-1,0)}^2 =0~,\\
\bar D_{(0,-1)}& :=& \bar D^\ad \frac{\pa}{\pa \bar \z^\ad}~ \qquad 
\bar D_{(0,-1)}^2 =0 
~.
\eea
\end{subequations}

Using the notation introduced,
the transverse linear condition \eqref{4.2a}  turns into 
\bea
\bar D_{(0,-1)} J_{(s,s-1)} &=& 0~,  \label{4.5a}
\eea
while the longitudinal linear condition \eqref{4.2b} takes the form
\bea
\bar D_{(0,1)} T_{(s-1,s-1)} &=& 0~. \label{4.5b}
\eea
The conservation equation \eqref{4.2c} becomes
\bea
-\frac{1}{2s} D_{(-1,0)} J_{(s,s-1)} + S_{(s-1,s-1)} + \bar T_{(s-1,s-1)} = 0~,
\label{4.6}
\eea
and \eqref{4.4} takes the form
\bea
\frac{1}{2s} D_{(1,0)} \left\{D_{(-1,0)} J_{(s,s-1)} + \bar D_{(0,-1)} \bar J_{(s-1,s)}\right\}
+D_{(1,0)} T_{(s-1,s-1)} =0~. 
\label{4.7}
\eea


\section{Higher spin supercurrents in a massive chiral model}

Consider the Fayet-Sohnius model \cite{Fayet,Sohnius}
for a free massive hypermultiplet
\bea
S_{\rm massive} = \int \rd^4x \rd^2 \q  \rd^2 \bar \q \, \Big( \bar \F_+ \F_+
+\bar \F_- \F_-\Big)
+\Big\{ {m} \int \rd^4x \rd^2 \q\, \F_+ \F_- +{\rm c.c.} \Big\}~,
\label{5.1}
\eea
where the superfields $\F_\pm$ are chiral, $\bar D_\ad \F_\pm =0$,
and  the mass parameter $m$ is chosen to be positive.

In the massless case, $m=0$, 
the conserved fermionic supercurrents $J_{\a(s) \ad(s-1)}$ 
were constructed in 
\cite{KMT}. In our notation they read
\bea
J_{(s,s-1)} &=& \sum_{k=0}^{s-1} (-1)^k
\binom{s-1}{k}
\left\{ \binom{s}{k+1} 
{\pa}^k_{(1,1)}
 D_{(1,0)} \F_{+} \,\,
{\pa}^{s-k-1}_{(1,1)}
 \F_{-}  
\right. \non \\ 
&& \left.
 \qquad \qquad
- \binom{s}{k} 
{\pa}^k_{(1,1)}
  \F_{+} \,\,
{\pa}^{s-k-1}_{(1,1)}
D_{(1,0)} \F_{-} \right\}~.
\label{4.8}
\eea
Making use of
the massless equations of motion,  $D^2 \F_\pm = 0$, 
one may check that $J_{(s,s-1)}$ obeys, 
for $s > 1$,  the conservation equations
\bea
D_{(-1,0)} J_{(s,s-1)} = 0, \qquad
\bar D_{(0,-1)} J_{(s,s-1)} = 0 ~.~
\label{4.9}
\eea

We will now construct fermionic higher spin supercurrents 
corresponding to  the massive model \eqref{5.1}.
Assuming that $J_{(s,s-1)}$ has the same functional form as in the massless case, 
eq. \eqref{4.8},  and making use of the equations of motion 
\bea
-\frac{1}{4} D^2 \F_+ +m \bar \F_{-} =0, \qquad
-\frac{1}{4} D^2 \F_- +m \bar \F_{+} =0,
\eea
we obtain 
\bea
D_{(-1,0)} J_{(s,s-1)} &=& 2m (s+1) \sum_{k=0}^{s-1} (-1)^{k+1} \binom{s-1}{k} \binom{s}{k} \non \\
&&\qquad \times \left\{ -\frac{s-k}{k+1} {\pa}^k_{(1,1)} \bar \F_- \,{\pa}^{s-k-1}_{(1,1)} \F_-
+ {\pa}^k_{(1,1)} \F_+ \,{\pa}^{s-k-1}_{(1,1)} \bar \F_+ \right\}\non \\
&&+ 2m(s+1) \sum_{k=1}^{s-1} (-1)^{k+1} \binom{s-1}{k} \binom{s}{k} \frac{k}{k+1} 
\non \\
&& \qquad \times {\pa}^{k-1}_{(1,1)} \bar D_{(0,1)} \bar \F_- \,{\pa}^{s-k-1}_{(1,1)} D_{(1,0)} \F_- 
\non \\
&&+ 2m(s+1)\sum_{k=0}^{s-2} (-1)^{k+1} \binom{s-1}{k} \binom{s}{k} \frac{s-1-k}{k+1} 
\non \\
&& \qquad \times {\pa}^{k}_{(1,1)} D_{(1,0)} \F_+ \,{\pa}^{s-k-2}_{(1,1)} \bar D_{(0,1)} \bar \F_+ ~.~ \label{4.10}
\eea
It can be shown that the massive supercurrent $J_{(s,s-1)}$ also obeys \eqref{4.5a}. 

We now look for a superfield $T_{(s-1,s-1)}$ such that (i) it obeys the longitudinal linear constraint \eqref{4.5b}; and (ii) it satisfies \eqref{4.7}, which is a consequence of the conservation equation \eqref{4.6}. 
We consider a general ansatz 
\bea
T_{(s-1, s-1)} &=& 
\sum_{k=0}^{s-1} c_k {\pa}^k_{(1,1)} \F_-\, {\pa}^{s-k-1}_{(1,1)} \bar \F_-  \non \\
&&+ \sum_{k=0}^{s-1} d_k {\pa}^k_{(1,1)} \F_+ \,
{\pa}^{s-k-1}_{(1,1)} \bar \F_+  \non \\
&&+ \sum_{k=1}^{s-1} f_k {\pa}^{k-1}_{(1,1)} D_{(1,0)} \F_-\, {\pa}^{s-k-1}_{(1,1)} \bar D_{(0,1)} \bar \F_-  \non \\
&&+ \sum_{k=1}^{s-1} g_k {\pa}^{k-1}_{(1,1)}  D_{(1,0)} \F_+\,\, {\pa}^{s-k-1}_{(1,1)} \bar D_{(0,1)} \bar \F_+ ~.
\label{T4.11}
\eea
Condition (i) implies that the coefficients must be related by
\begin{subequations} 
\bea
c_0 = d_0 = 0~, \qquad f_k = c_k~, \qquad g_k = d_k~, 
\eea
while for $k=1,2, \dots s-2$, condition  (ii) gives the following recurrence relations:
\bea 
c_k + c_{k+1} &=& 
\frac{m(s+1)}{s} (-1)^{s+k} \binom{s-1}{k} \binom{s}{k} 
\non \\
&& \times \frac{1}{(k+2)(k+1)} \Big\{(2k+2-s)(s+1)-k-2\Big\}~, \\
d_k + d_{k+1} &=& \frac{m(s+1)}{s} (-1)^{k} \binom{s-1}{k} \binom{s}{k} 
\non \\
&& \times \frac{1}{(k+2)(k+1)} \Big\{(2k+2-s)(s+1)-k-2\Big\}~.
\eea
Condition (ii) also implies that
\bea
c_1 = -(-1)^s \frac{m(s^2-1)}{2}~, \qquad c_{s-1} &=&- \frac{m(s^2-1)}{s}~;\\
d_1 =- \frac{m(s^2-1)}{2}~, \qquad d_{s-1} &=& -(-1)^s \frac{m(s^2-1)}{s}~.
\eea
\end{subequations}
The above conditions lead to simple expressions for $c_k$ and $d_k$:
\begin{subequations}
\bea
d_k &=& \frac{m(s+1)}{s} \frac{k}{k+1} (-1)^{k} \binom{s-1}{k} \binom{s}{k}~,\\
c_k &=& (-1)^s d_k ~,
\eea
\end{subequations}
where $ k=1,2,\dots s-1$.
Now that we have already derived an expression for the trace multiplet $T_{(s-1,s-1)}$, the superfield $S_{(s-1,s-1)}$ can be computed using the conservation equation \eqref{4.6}. This gives
\bea
S_{(s-1,s-1)} &=& -m (s+1)\sum_{k=0}^{s-1} (-1)^{k+1}
\binom{s-1}{k}\binom{s}{k} \frac{1}{k+1}
 \non \\
 && \qquad \qquad
\times \left\{ 
{\pa}^k_{(1,1)}
\bar \F_{-} \,
{\pa}^{s-k-1}_{(1,1)}
 \F_{-}  
+ (-1)^s  
{\pa}^k_{(1,1)}
 \bar \F_{+} \,
{\pa}^{s-k-1}_{(1,1)}
 \F_{+} \right\}~.
\eea
One may verify that $S_{(s-1,s-1)}$ is a real superfield. 

\section{Concluding comments}

To conclude this work, we make several final comments. 

The formulation proposed in section \ref{section2} can naturally 
be lifted to the case of anti-de Sitter supersymmetry to extend
the results of \cite{KS94}.

The action \eqref{action} involves the transverse linear compensator 
$\S_{\a(s-1) \ad (s-2) }$ and its conjugate $\bar \S_{\a (s-2) \ad(s-1)}$. 
These superfields cannot be dualised into a longitudinal linear supermultiplet
without destroying the locality of the theory, for the action  \eqref{action} 
contains terms with derivatives of 
$\S_{\a(s-1) \ad (s-2) }$ and $\bar \S_{\a (s-2) \ad(s-1)}$. 

The hypermultiplet model is $\cN=2$ supersymmetric, and therefore its 
conserved currents should belong  to $\cN=2$ supermultiplets. 
In the massless case, $m=0$, we deal with the $\cN=2$ Poincar\'e 
supersymmetry without central charge on the mass shell. 
In this case it  is easy to embed
the bosonic $J_{\a(s) \ad(s)}$ and fermionic $J_{\a(s) \ad(s-1)}$ higher spin supercurrents, which were constructed in \cite{KMT} for any $s\geq1$, 
into $\cN=2$ 
real superfields ${\mathbb J}_{\a(s-1) \ad(s-1)}= \bar {\mathbb J}_{\a(s-1) \ad(s-1)}$   
introduced in \cite{HST} and constrained by 
\bea
D_i^\b {\mathbb J}_{\b \a(s-2)  \ad(s-1)}=0 \quad \Longleftrightarrow \quad
\bar D_i^\bd {\mathbb J}_{\a(s-1) \bd \ad(s-2)}=0~, \qquad i =\1, \2~.
\eea
Here $D^i_\a$ and $\bar D_i^\ad$ are the spinor covariant derivatives 
of $\cN=2$ Minkowski superspace. 
Conserved
$\cN=1$ supercurrent multiplets originate as 
\begin{subequations}\label{6.2}
\bea
{J}_{\a(s-1) \ad(s-1)}&:=& {\mathbb J}_{\a(s-1) \ad(s-1)}|~, 
\\
{J}_{\a(s) \ad(s-1)}&:=& D^\2_{\a_1}{\mathbb J}_{\a_2 \dots \a_s \ad(s-1)}|~,\\
{J}_{\a(s) \ad(s)}&:=& \hf \Big( \big[D^\2_{(\a_1}, \bar D_{\2 (\ad_1} \big]
-\frac{1}{2s+1} \big[D^\1_{(\a_1}, \bar D_{\1 (\ad_1} \big] \Big)
{\mathbb J}_{\a_2 \dots \a_s) \ad_2 \dots \ad_s)}|
~,
\eea
\end{subequations}
where we have made use of the $\cN=1$ projection, 
$U| := U(x,\q^\a_i, {\bar \q}^j_\ad ) 
|_{ \q_{\underline{2}} = 
{\bar \q}^{\underline{2}} = 0}$, of any $\cN=2$ superfield
$U$.\footnote{In this setting, the $\cN=1$ spinor covariant derivatives 
are identified as 
$D_\a := D^{\underline{1}}_\a$ and  ${\bar D}^\ad :=
{\bar D}^\ad_{\underline{1}}$.}  
In the $s=1$ case, the relations \eqref{6.2} reduce to those in eq. (1.10)
of \cite{KT}.

In the massive case, $m\neq0$, we deal with the $\cN=2$ Poincar\'e 
supersymmetry with a constant central charge on the mass shell, 
and the story becomes pretty subtle. 
In our previous work \cite{HK}, we observed that the higher spin supercurrents $J_{\a(s)\ad(s)}$ in the massive chiral model exist only for odd values of $s$. 
The same conclusion was also reached in  a revised version 
(v3, 26 Oct.) of Ref. \cite{BGK}.
However, the conserved fermionic supercurrents $J_{\a(s)\ad(s-1)}$ 
constructed in the present paper
are realised for all values of $s>1$.

The longitudinal and transverse actions for the 
massless integer superspin-$s$ multiplet \cite{KS} are well defined 
for $s=1$, in which case they describe two off-shell formulations for the 
massless gravitino multiplet. However, the action \eqref{action} 
is not defined in the $s=1$ case. The point is that 
the gauge transformation law \eqref{2.5a} is not defined for $s=1$.
The gauge freedom in the superconformal gravitino multiplet model \cite{KMT} is 
\begin{subequations}
\bea
\d \J_\a = \hf D_\a {\frak V} + \z_\a ~, \qquad \bar D_\bd \z_\a =0~.
\label{6.3a}
\eea
This transformation law of $\J_\a$ coincides with the one occurring in the 
off-shell model for the massless gravitino multiplet proposed in \cite{GS}.
In addition to the gauge superfield $\J_\a$, 
this model also involves two compensators, a real scalar $H$ and a chiral scalar $\F$, 
$\bar D_\ad \F=0$, with the gauge transformation laws
\bea 
\d H &=& {\frak V} + \bar {\frak V} ~, \label{6.3b}\\
\d \F&= & -\hf \bar D^2 \bar {\frak V} ~.
\eea
\end{subequations}
The gauge invariant action of \cite{GS} can be written in the form \cite{Ideas}
\bea
S^{\rm (I)}_{\rm GM}= 
S^{\|}_{(1, \frac 32 )} [\J, \bar \J, H] - \hf \int  \rd^4x \rd^2 \q  \rd^2 \bar \q\,
\Big( \bar \F \F +\F D^\a \J_\a +\bar \F \bar D_\ad \bar \J^\ad \Big)~,
\label{6.4}
\eea
where $S^{\|}_{(1, \frac 32 )} [\J, \bar \J, H] $ denotes the longitudinal action for the gravitino multiplet, 
which is obtained from \eqref{action} by choosing the gauge \eqref{2.9}
and setting $s=1$. At the component level, this manifestly supersymmetric model is known to describe
the Fradkin-Vasiliev-de Wit-van Holten formulation for the gravitino multiplet
\cite{FV,deWvanH}.

There exists a dual formulation  for \eqref{6.4} that is  obtained by performing a superfield Legendre transformation \cite{LR2}. The dual action given in \cite{LR2} is 
\bea
S^{\rm (II)}_{\rm GM}= 
S^{\|}_{(1, \frac 32 )} [\J, \bar \J, H] +\frac 14 \int  \rd^4x \rd^2 \q  \rd^2 \bar \q\,
\Big( G + D^\a \J_\a + \bar D_\ad \bar \J^\ad \Big)^2~,
\label{6.5}
\eea
where $G=\bar G$ is a real linear superfield, $\bar D^2 G = D^2G=0$.
The gauge freedom in this theory is given by eqs. \eqref{6.3a}, \eqref{6.3b}
and 
\bea
\d G = -D^\a \z_\a -\bar D_\ad \bar \z^\ad ~,
\eea
in accordance with \cite{ButterK}.
It may be used 
to impose two conditions $H =0$ and $G=0$. Then we end up 
with the Ogievetsky-Sokatchev formulation for the gravitino multiplet 
\cite{OS} (see section 6.9.5 \cite{Ideas} for the technical details).

Actually, there exists one more dual formulation  for \eqref{6.4} that is  obtained
by performing the complex linear-chiral duality transformation. 
 It leads to 
\bea
S^{\rm (III)}_{\rm GM}= 
S^{\|}_{(1, \frac 32 )} [\J, \bar \J, H] +\frac 12 \int  \rd^4x \rd^2 \q  \rd^2 \bar \q\,
( \S + D^\a \J_\a)(\bar \S  + \bar D_\ad \bar \J^\ad )~,
\label{6.6}
\eea
where $\S$ is a complex linear superfield  constrained by $\bar D^2 \S=0$. 
The gauge freedom in this theory is given by eqs. \eqref{6.3a}, \eqref{6.3b}
and 
\bea
\d \S = -D^\a \z_\a  ~.
\eea
This gauge freedom does not allow one to gauge away $\S$  off the mass shell. 
To the best of our knowledge, the supersymmetric gauge theory \eqref{6.6} is a
new off-shell realisation for the massless gravitino multiplet. 

As shown in \cite{ButterK}, the gravitino multiplet actions \eqref{6.4} and \eqref{6.5}
naturally originate upon $\cN=2 \to \cN=1$ reduction of the linearised superfield action 
\cite{ButterK} for the off-shell $\cN=2$ supergravity with a tensor compensator
\cite{deWPV}. The actions \eqref{6.4} and \eqref{6.5} prove to correspond to different values of the background tensor multiplet \cite{ButterK}. 
The gravitino multiplet action \eqref{6.6} should originate if one linearises 
the off-shell $\cN=2$ supergravity with a tropical  compensator
\cite{KLRT-M}. 

The transverse formulation for the massless gravitino multiplet, which was introduced 
in \cite{KS}, is quite mysterious in the sense that it is not contained in any known 
off-shell formulation for $\cN=2$ supergravity.
\\

\noindent
{\bf Acknowledgements:}\\
The work of JH is supported by an Australian Government Research Training Program (RTP) Scholarship.
The work of SMK is supported in part by the Australian 
Research Council, project No. DP160103633.


\begin{footnotesize}

\end{footnotesize}

\end{document}